\documentclass{article}

\usepackage{spconf,amsmath,graphicx,hyperref}

\usepackage[caption=false,font=normalsize,labelfont=sf,textfont=sf]{subfig}
\usepackage{textcomp}
\usepackage{stfloats}
\usepackage{url}
\usepackage{verbatim}
\usepackage{cite}
\usepackage{cases}

\usepackage{amsfonts,amssymb} 
\usepackage{array}                    
\usepackage{bm}                       
\usepackage{algorithmic, algorithm} 
\usepackage{amsthm}                 

\usepackage{booktabs}
\usepackage{threeparttable}

\usepackage{color}

\def\BibTeX{{\rm B\kern-.05em{\sc i\kern-.025em b}\kern-.08em
    T\kern-.1667em\lower.7ex\hbox{E}\kern-.125emX}}
\definecolor{orange}{RGB}{255,107,0}

\definecolor{green}{RGB}{0,180,80}

\newcommand\diag{\ensuremath{{\rm diag}}}

\newcommand\st{\ensuremath{{\rm s.t.}}}

\newtheorem{Lemma}{Lemma}

\newcommand\Ch{\ensuremath{\mathbb{C}}}

\newcommand\Cc{\ensuremath{\mathcal{C}}}

\newcommand\Kc{\ensuremath{\mathcal{K}}}    

\newcommand\Nc{\ensuremath{\mathcal{N}}}
\newcommand\Oc{\ensuremath{\mathcal{O}}}

\newcommand\Qc{\ensuremath{\mathcal{Q}}}

\newcommand\Tc{\ensuremath{\mathcal{T}}}	

\def\A {\mathbf A}
\def\a {\mathbf a}	

\def\tbA{\tilde{\A}}

\def\b {\mathbf b}

\def\c {\mathbf c}	

\def\G {\mathbf G}
\def\g {\mathbf g}

\def\tbG{\tilde{\G}}	

\def\H {\mathbf H}	
\def\h {\mathbf h}	

\def\I {\mathbf I}

\def\q {\mathbf q}

\def\w {\mathbf w}	  

\def\y {\mathbf y}

\def\bPhi {\boldsymbol \Phi}

\def\bphi {\boldsymbol \phi}

\newcommand\Ebb{\ensuremath{{\mathbb{E}}}}
\def\diag{\mathrm{diag}}

\title{RIS-Enhanced Information-Decoupled Symbiotic Radio Over Broadcasting Signals
\thanks{Accepted to ICASSP 2026. \copyright~2026 IEEE. Author's version.
}}

\name{Shu Cai$^{\star}$ \qquad Ya-Feng Liu$^{\dagger}$ \qquad Jun Zhang$^{\star}$ \qquad Qi Zhang$^{\star}$}
\address{ $^{\star}$College of Telecommunications and Information Engineering, \\ Nanjing University of Posts and Telecommunications, Nanjing, China\\ 
$^{\dagger}$School of Mathematical Sciences, Beijing University of Posts and Telecommunications, Beijing, China}

\begin{document}
%
\maketitle
\begin{abstract}
This paper studies a reconfigurable intelligent surface (RIS)-enhanced decoupled symbiotic radio (SR) system in which a  primary transmitter delivers common data to multiple primary receivers (PRs), while a RIS-based backscatter device sends secondary data to a backscatter receiver (BRx). Unlike conventional SR, the BRx performs energy detection and never decodes the primary signal, thereby removing ambiguity and preventing exposure of the primary payload to unintended receivers. In this paper, we formulate the problem as the minimization of the transmit power subject to a common broadcast rate constraint across all PRs and a bit error rate (BER) constraint at the BRx. The problem is nonconvex due to the unit-modulus RIS constraint and coupled quadratic forms. Leveraging a rate-balanced reformulation and a monotonic BER ratio characterization, we develop a low-complexity penalty-based block coordinate descent algorithm with closed-form updates. Numerical results show fast convergence of the proposed algorithm and reduced power consumption of the considered RIS-enhanced information-decoupled SR system over conventional SR baselines.
\end{abstract}

\section{Introduction}
The vision for 6G wireless networks anticipates massive connectivity, low-power sensing, and enhanced spectrum reuse for IoT applications \cite{6GIoTWC,6GBell,6GIoTCC}. Ambient backscatter communication (AmBC) has emerged as a promising paradigm for ultra-low-power links by leveraging existing radio-frequency (RF) carriers; however, its performance is fundamentally constrained by weak backscattered signals and dependence on ambient sources, which severely limit coverage and reliability \cite{AmBC2013ACM,Liang2022Backscatter,Gao2017TC}.

Symbiotic radio (SR) addresses these limitations by allowing an existing legacy primary transmission to serve as a host signal for a secondary backscatter device (BDx). The BDx modulates reflections to transmit its own data, while potentially providing diversity or coding benefits to the primary link via joint reception at the primary receiver (PR). Recent studies have developed the core principles and variants of SR, highlighting its potential for green spectrum-efficient IoT connectivity \cite{Liang2020TCCN,Shiying2021TWC,Long2020IoT}. When combined with reconfigurable intelligent surfaces (RISs), which can shape propagation via nearly-passive phase control, the composite channel can be enhanced for both primary and secondary links, enabling coverage extension and interference suppression in a hardware-affordable manner \cite{PanRIS2022,zhou_cooperative_2022,ZQQ2021TC,Xu2021TC,Hu2021TC,JinShiRIS2023,Vedio2025,ARIS2025}.

Despite these advances, practical deployments of SR confront two challenges at the backscatter receiver (BRx): (i) persistent bit error rate (BER) floors when the direct primary link to the BRx is weak; and (ii) decoding ambiguity when the direct link is blocked. Both of the above challenges stem from the requirement that the BRx must decode the primary signal \cite{Zhou2024TCCN,Zhou2025TWC}. Moreover, exposing the primary payload to the BRx raises intrinsic confidentiality concerns \cite{ShuRISSRTVT2025}. To reconcile energy efficiency, reliability, and confidentiality, we consider an information-decoupled SR (ID-SR) design in which the BRx performs energy detection and never decodes the primary signal, while a single PR executes SR-style joint decoding to exploit the backscatter-induced structure \cite{ShuRISSRTVT2025}.

Many practical services, such as firmware broadcasting, command-and-control updates, and public-safety alerts, involve disseminating a common message to multiple PRs. At the same time, a RIS-based BDx (RIS-BDx) may send a low-rate sensing or control stream to a separate BRx. Extending ID-SR from a single PR to a broadcast setting introduces new technical challenges. For instance, the primary quality of service becomes the common rate across all PRs, which couples heterogeneous channels through a single transmit beamformer and a shared RIS phase profile.

This paper addresses these challenges by formulating a broadcast ID-SR design problem in which a multi-antenna primary transmitter (PTx) delivers a common primary stream to multiple PRs via joint decoding, while a RIS-BDx modulates a secondary symbol that the BRx detects through energy detection. We develop a penalty-based block-coordinate descent (PBCD) method that converts the nonconvex constraints into smooth surrogate penalties, yielding closed-form updates for the primary beamformer and RIS phases. The key feature of the proposed algorithm is that it scales linearly with the number of RIS elements. Numerical results show fast convergence of the proposed algorithm and reduced power consumption of the considered RIS-enhanced ID-SR system over conventional SR baselines.

\section{System Model and Problem Formulation}\label{sec:sys}
\subsection{Network and Channel Model}
\begin{figure}[!t]
	\begin{center}
		{\resizebox{0.4\textwidth}{!}
			{\includegraphics{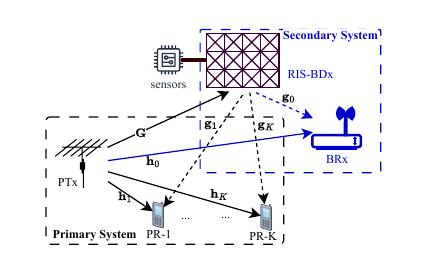}}}
	\end{center}\vspace{-0.5cm}
	\caption{The RIS-Enhanced ID-SR (RIS-ID-SR) over broadcasting signals.}\label{fig:sysmod}
	\vspace{-0.3cm}
\end{figure}

We consider a RIS-enhanced SR downlink system consisting of one $N_t$-antenna PTx, $K$ single-antenna PRs, a RIS-BDx with $N_r$ reflecting elements, and a single-antenna BRx; see Fig.~\ref{fig:sysmod}. 
The RIS applies a unit-modulus diagonal matrix $\bPhi=\diag(\bphi)$ with
{\small
\begin{align}\label{eq:phi}
\bphi=[\phi_1,\phi_2,\ldots,\phi_{N_r}]^\top  \text{ with } \phi_n=e^{j\theta_n},~\forall n.
\end{align}}%
The PTx broadcasts a common primary stream to all PRs while the BRx receives an additional secondary stream embedded via RIS backscatter. The RIS-BDx modulates its own symbols to the BRx by imposing symbol-dependent phase changes on the incident waveform.

Let $\h_k\in\Ch^{N_t\times 1}$ and $\h_0\in\Ch^{N_t\times 1}$ denote the channels from PTx to PR-$k$ and from PTx to BRx, respectively. Let $\G\in\Ch^{N_t\times N_r}$ be the channel from PTx to RIS. Also, let  $\g_k\in\Ch^{N_r\times 1}$ and $\g_0\in\Ch^{N_r\times 1}$ denote the channels form RIS to PR-$k$ and from RIS to BRx, respectively. We assume perfect cascaded channel state information (CSI) at the PTx to characterize the performance limits of the considered system. Practical CSI acquisition can follow protocols such as \cite{ZQQ2021TC}.

\subsection{Signal Model}
\begin{figure}[t]
	\begin{center}
		{\resizebox{0.42\textwidth}{!}
			{\includegraphics{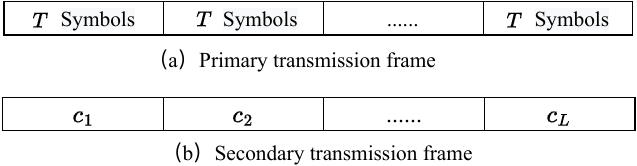}}}
	\end{center}\vspace{-0.5cm}
  \caption{Symbols in the RIS-ID-SR system, where one secondary symbol period comprises $T$ primary symbols. Due to the weak power of the backscattered signal, the achievable rate at the BRx is lower than that at the PRx.}\label{fig:frames}
	\vspace{-0.3cm}
\end{figure}

The system operates over coherence blocks of $L T$ primary symbols. 
During the $\ell$-th secondary symbol period, the RIS-BDx emits a binary symbol $c_\ell\in\Qc_c\triangleq\{\pm1\}$ while the PTx continuously transmits the primary signal $s_\ell(t)$ over $T$ primary symbol slots, as illustrated in Fig.~\ref{fig:frames}. The received signals at PR-$k$ and the BRx are
{\small
\begin{subequations}\label{eq:rx}
\begin{align}
y_{k,\ell}(t)=&(\h_k +\G \bPhi\g_k{c_{\ell}})^H \w{s_{\ell}(t)}+n_{k,\ell}(t), ~k\in \Kc,\label{eq:rx_pr}\\
y_{0,\ell}(t)=&(\h_0 +\G \bPhi\g_0{c_{\ell}})^H \w{s_{\ell}(t)}+n_{0,\ell}(t),\label{eq:rx_b}
\end{align}
\end{subequations}}%
where $\w$ is the PTx precoder, $n_{k,\ell}\sim\Cc\Nc(0,\delta^2)$ and $n_{0,\ell}\sim\Cc\Nc(0,\delta^2)$ are AWGN, and $\Kc=\{1,2,\ldots,K\}$. 

\subsection{Primary Rates and BRx Energy Detection}
For $c_\ell\!\in\!\Qc_c$, define the effective channels
{\small
\begin{align}
\h_{E,k}(i) = \h_k - (-1)^i\G \bPhi\g_k, ~i\in\{0,1\}.\label{eq:effch}
\end{align}}%
Assume $\Ebb\{|s_\ell(t)|^2\}=1$. Then the achievable rate at PR-$k$ is
{\small
\begin{align}
R_k=\Ebb_{c_\ell}\left[\log_2\left(1+{|\h_{E,k}^H(c_\ell)\w|^2}/{\delta^2}\right)\right],~k\in\Kc. \label{eq:ratek}
\end{align}}%

For the BRx, we approximate $s_\ell(t)$ as i.i.d. $\Cc\Nc(0,1)$ \cite{Gao2017TC}, yielding $y_{0,\ell}(t)\sim \Cc\Nc(0,\sigma_i^2)$ under $c_\ell=(-1)^i$, where
{\small
\begin{align}
\sigma_i^2=|\h_{E,0}^H(i)\w|^2+\delta^2,\quad i\in\{0,1\}. \label{eq:sigma}
\end{align}}%
The corresponding likelihood-ratio test is an energy threshold on $\| \y_{0,\ell}\|^2\triangleq \sum_{t\in\Tc}|y_{0,\ell}(t)|^2$ with the closed-form BER \cite{ShuRISSRTVT2025}
{\small
\begin{align}
P_b(\tfrac{\sigma_1^2}{\sigma_0^2},T)=\frac{1}{2\Gamma(T)}\!\left[\gamma\!\Big(T,\!T\sigma_{\min}^2\Delta\Big)\!+\!\Gamma\!\Big(T,\!T\sigma_{\max}^2\Delta\Big)\right], \label{eq:ber}
\end{align}}%
where $\sigma_{\min}^2=\min(\sigma_0^2,\sigma_1^2)$, $\sigma_{\max}^2=\max(\sigma_0^2,\sigma_1^2)$, $\Delta\!= \!\tfrac{\ln(\sigma_1^2)-\ln(\sigma_0^2)}{\sigma_1^2-\sigma_0^2}$, and $\Gamma(T)$, $\gamma(T, x)$ and $\Gamma(T, x)$ denote the gamma, lower incomplete gamma, and upper incomplete gamma functions, respectively. 

\subsection{Transmit Power Minimization}\label{sec:form}
Our goal is to jointly design $\w$ and $\bphi$ to minimize the transmit power, subject to the following two constraints: (i) each PR achieves the broadcast rate not less than the target $r_p$ and (ii) the BRx's BER does not exceed $\Lambda_s$. Formally, we require $R_k \ge r_p$ for all $k\in\Kc$ and $P_b(\tfrac{\sigma_1^2}{\sigma_0^2},T) \le \Lambda_s$. These constraints are nonconvex and challenging to handle directly.

Following \cite{ShuRISSRTVT2025}, we adopt two reformulations: 
\textbf{(i) Rate-balanced constraints.}  To avoid possible primary rate splitting due to stochastic $c_{\ell}$ in \eqref{eq:ratek}, we approximate the rate constraint $R_k\geq r_p$ by the following inequalities:
{\small
\begin{align}
|\h_{E,k}^H(i)\w|^2 \ge \delta^2 \Gamma_p,\ \ \Gamma_p\triangleq 2^{r_p}-1,\ \ i\in\{0,1\},~k\in\Kc. \label{eq:rb}
\end{align}}%
This ensures that the primary rate remains consistent for both RIS states.
\textbf{(ii) BER ratio form.} When $\sigma_1^2\ge \sigma_0^2$, $P_b(\tfrac{\sigma_1^2}{\sigma_0^2},T)$ is monotonically decreasing in $\tfrac{\sigma_1^2}{\sigma_0^2}$. Thus, $P_b\le \Lambda_s$ is equivalent to
 $|\h_{E,0}^H(1)\w|^2 - \lambda_s |\h_{E,0}^H(0)\w|^2 + (1-\lambda_s)\delta^2 \ge 0$, 
where $\lambda_s\geq 1$ satisfies $P_b(\lambda_s,T)=\Lambda_s$.
Then, we obtain the transmit power minimization problem as follows:
{\small
\begin{subequations}\label{eq:P1}
\begin{align}
\min_{\w,\bphi}\ \ & \|\w\|^2 \\
\st\ \ & |\h_{E,k}^H(i)\w|^2 \ge \delta^2\Gamma_p,\ \ i\in\{0,1\},\ k\in\Kc, \label{eq:P1_rate}\\
& |\h_{E,0}^H(1)\w|^2 \!-\! \lambda_s |\h_{E,0}^H(0)\w|^2 \!+\! (1-\lambda_s)\delta^2 \ge 0, \label{eq:P1_ber}\\
& |\phi_n|=1,\ \forall n. \label{eq:P1_cm}
\end{align}
\end{subequations}}%

The BER constraint \eqref{eq:P1_ber} admits a companion inequality when $\sigma_0^2\ge \sigma_1^2$. However, it is sufficient to solve only one branch. With BPSK backscatter $c_\ell\in\{\pm1\}$, replacing $\bPhi$ with $-\bPhi$ swaps the effective channels as $\h_{E,k}(1)\leftrightarrow \h_{E,k}(0)$ for all $k=0,1,\dots,K$. Consequently, if $(\w,\bphi)$ solves one branch of \eqref{eq:P1}, then $(\w,-\bphi)$ solves the other with the same transmit power. Compared with \cite{ShuRISSRTVT2025}, problem \eqref{eq:P1} includes coupled rate constraints for multiple PRs and a BER constraint at the BRx, which renders the techniques in \cite{ShuRISSRTVT2025} inapplicable. In the following, we shall develop a new approach to solving problem \eqref{eq:P1}.

\section{The Proposed PBCD Method}\label{sec:method}
Problem \eqref{eq:P1} contains quartic terms in $(\w,\bphi)$ as they are coupled in the rate and BER constraints. We therefore adopt a PBCD framework with auxiliary variables to decouple constraints, yielding simple updates for $\w$, $\bphi$, and the auxiliaries.

\subsection{Penalty Reformulation}
Introduce the effective-channel stacks $\H_{E,i}\triangleq [\h_{E,0}(i),$  $\h_{E,1}(i),\dots,\h_{E,K}(i)]$ and auxiliary vectors $\q_i\triangleq \H_{E,i}^H\w$ for $i\in\{0,1\}$, and consider the quadratic penalty functions:
{\small
\begin{subequations}\label{eq:penalty}
\begin{align}
\min_{\w,\bphi,\q_0,\q_1}\ \ & \|\w\|^2 + \rho\sum_{i=0}^1 \|\q_i-\H_{E,i}^H\w\|^2 \label{eq:penalty_obj}\\
\st\quad~  & |q_{k,i}|^2 \ge \delta^2\Gamma_p,\ \ i=0,1,\ k\in\Kc, \label{eq:penalty_rate}\\
& q_{0,1}^2 \ge \lambda_s |q_{0,0}|^2 + (\lambda_s-1)\delta^2,\ \Im\{q_{0,1}\}=0, \label{eq:penalty_ber}\\
& |\phi_n|=1,\ \forall n, \label{eq:penalty_cm}
\end{align}
\end{subequations}}%
where $q_{k,i} = \h_{E,k}^H(i)\w$ and $\rho>0$ is the penalty parameter. Without loss of optimality one can rotate $\w$ so that $q_{0,1}$ is real. 
We apply BCD to solve problem \eqref{eq:penalty} and increase $\rho$ geometrically until the constraint residuals $\|\q_i-\H_{E,i}^H\w\|_{\infty}$ are below the desired tolerance.

\subsection{
The \texorpdfstring{$q_{k,i}$}{q\_{k,i}}-subprobelms with \texorpdfstring{$k\geq 1$}{k>=1}} 

Given $(\w,\bphi)$, each $q_{k,i}$ solves a projection:
{\small
\begin{align}\label{eq:qki_sol}
q_{k,i}^{\star}= \max\!\left\{1,{\sqrt{\Gamma_p}\delta}/{|\h_{E,k}^H(i)\w|}\right\}\h_{E,k}^H(i)\w.
\end{align}}

\subsection{The \texorpdfstring{$(q_{0,0},q_{0,1})$}{q\_{0,i}}-subproblem}

With $t_0 \triangleq \h_{E,0}^H(0)\w$ and $t_1 \triangleq \Re\{\h_{E,0}^H(1)\w\}$, we solve
{\small
\begin{subequations}\label{eq:q00q01}
\begin{align}
\min_{q_{0,0},\,q_{0,1}}\ \ &(q_{0,1}-t_1)^2 + |q_{0,0}-t_0|^2 \\
\st\quad~ & q_{0,1}^2 \ge \lambda_s |q_{0,0}|^2 + (\lambda_s-1)\delta^2. \label{eq:q00q01_c}
\end{align}
\end{subequations}}%
To simplify problem \eqref{eq:q00q01}, we induce the following lemma. 
\begin{Lemma}\label{lem:ti}
  The optimal solutions of \eqref{eq:q00q01} satisfy 
{\small
\begin{align}\label{eq:q00q01_sol}
q_{0,0}^{\star}=c_0 t_0,\quad q_{0,1}^{\star}=c_1 t_1,\ \ c_0,c_1\ge 0.
\end{align}}
\end{Lemma}
Lemma \ref{lem:ti} can be proved by contradiction; however, its proof is omitted for brevity. 
Substituting \eqref{eq:q00q01_sol} into problem \eqref{eq:q00q01} leads to
{\small \begin{subequations}\label{eq:c00c01}
	\begin{align}
		\min_{c_0>0,c_1>0}~ &  (c_0 - 1)^2|t_0|^2 + (c_1 - 1)^2 t_1^2 \label{eqBCP1S2F3TS1:obj}\\
		\st
          ~~~~~& c_1^2 t_1^2 \geq \lambda_s c_0^2|t_0|^2+(\lambda_s-1)\delta^2.\label{eqBCP1S2F3TS1:Pb}
	\end{align}
\end{subequations}}%
By using the Lagrange multiplier method, it can be derived that 
{\small \begin{align}\label{eqBC:P1S2F3Tsol2}
c_0(\lambda_t) &=(1+\lambda_t \lambda_s)^{-1},~~
c_1(\lambda_t) = (1-\lambda_t)^{-1},
\end{align}}%
where $\lambda_t\geq 0$ is the Lagrange multiplier. Then, either $\lambda_t = 0$ or $\lambda_t$ is a root to the equation in \eqref{eqBCP1S2F3TS1:Pb}.

\subsection{The \texorpdfstring{$\w$}{w}-subproblem} 
For fixed $(\bphi,\q_0,\q_1)$, \eqref{eq:penalty_obj} is quadratic and strictly convex in terms of $\w$. Then we have
{\small
\begin{align}\label{eq:w_update}
\w^{\star}=\Big(\sum_{i=0}^1\H_{E,i}\H_{E,i}^H+\rho^{-1}\I\Big)^{-1}\Big(\sum_{i=0}^1\H_{E,i}\q_i\Big).
\end{align}}%

\subsection{The RIS Phase Update} 
For fixed $(\w,\q_0,\q_1)$, the $\bphi$-dependent terms in \eqref{eq:penalty} are
{\small 
	\begin{align}\label{eqBC:P1S2FPhi}
		\min_{\substack{|\phi_n|=1 \\ n=1,2,\dots,N_r}}~
  &\sum_{i=0}^1\sum_{k=0}^K\|q_{k,i} - (\h_k - (-1)^i\G \bPhi\g_k)^H\w\|^2.
	\end{align}}%
Using $\G \bPhi\g_k = \G\diag(\g_k)\bphi$ and denoting $\tbG_k = \G\diag(\g_k)$, we can 
rewrite \eqref{eqBC:P1S2FPhi} as 
{\small \begin{align}\label{eqBC:P1S2FPhi1}
\min_{\substack{|\phi_n|=1 \\ n=1,2,\dots,N_r}}~\|\A_{\phi}\bphi - \b_{\phi}\|^2, 
\end{align}}%
where $\b_{\phi} = [\b_{\phi,0}^{\top},\b_{\phi,1}^{\top}]^{\top}$, $\A_{\phi} = [\tbA_{\phi}^{\top},-\tbA_{\phi}^{\top}]^{\top}$ with 
{\small
\begin{subequations}\label{eqBC:phidefs}
	\begin{align}
         \b_{\phi,i} &= (\q_i^H - \w^H[\h_0,\h_1,\dots,\h_K])^{\top},\\
        \tbA_{\phi,i} &= [ (\w^H\tbG_0)^{\top},(\w^H\tbG_1)^{\top},\dots,(\w^H\tbG_K)^{\top}]^{\top}.
	\end{align}
\end{subequations}}%
Then element-wise updating of $\phi_n$ yields 
{\small \begin{align}\label{eq:optphi}
    \phi_n^{\star} = (\a_{\phi,n}^H\c_{\phi,n})/|\a_{\phi,n}^H\c_{\phi,n}|,
\end{align}}%
where $\a_{\phi,n}$ is the $n$-th column of $\A_{\phi}$, $\c_{\phi,n} = \b_{\phi} - \A_{\phi}\bphi_n$, and $\bphi_n$ denotes $\bphi$ with its $n$-th entry set to zero.

\subsection{Algorithm and Complexity}
The complete procedure of the proposed PBCD algorithm for solving problem \eqref{eq:P1} is summarized in Algorithm~\ref{alg:BCD}. The per-iteration complexity is dominated by inverting an $N_t\times N_t$ matrix in \eqref{eq:w_update} (which can be accelerated via the matrix inversion lemma), whereas each element-wise RIS phase update in \eqref{eq:optphi} requires $\Oc(1)$ operations, leading to $\Oc(N_r)$ complexity for a full sweep.

\begin{algorithm}[t]
\caption{BCD with Quadratic Penalties for \eqref{eq:P1}}\label{alg:BCD}
\begin{algorithmic}[1]
\STATE \textbf{Initialize:} $\w$, $\bphi$; set $\rho>0$, $c_{\rho}>1$, $\varepsilon_1>0$, $\varepsilon_2>0$.
\REPEAT
\STATE Set $\rho\leftarrow c_{\rho}\rho$.
\REPEAT
\STATE Update $(\q_0,\q_1)$ by \eqref{eq:qki_sol}--\eqref{eq:q00q01_sol}.
\STATE Update $\phi_n$, $n=1,2,\dots,N_r$, sequentially by \eqref{eq:optphi}.
\STATE Update $\w$ by \eqref{eq:w_update}.
\UNTIL {the relative change in norm of variables $<\varepsilon_1$.}
\UNTIL{the equality constraint violation $<\varepsilon_2$.}
\end{algorithmic}
\end{algorithm}

\section{Simulation Results}\label{sec:sims}
This section presents numerical results to demonstrate the effectiveness of the proposed RIS-ID-SR system and the PBCD algorithm. Simulations are conducted under standard large-scale path loss and small-scale Rayleigh fading \cite{ShuRISSRTVT2025}. 
A three-dimensional coordinate system as shown in Fig.~\ref{fig:sysmod} is considered, where $N_t = 16$, $N_r = 2^{10}$, $K = 4$, and $\Gamma_p = 15$ dB, $T = 50$, and $\Lambda_s=0.0786$ by default.

We compare the following systems: (1) the system without RIS and without BRx under a PR rate constraint (``WORIS''); (2) the system with RIS but without BRx (``WOBRx''); (3) a C-SR system where the BRx uses joint decoding \cite{Xu2021TC} with the joint design problem as follows:
\begin{subequations}\label{eq:PLiang}
	\begin{align}
		\min_{\w,\bphi}~&    \|\w\|^2  \\
		\st ~& R_{k}\geq r_p,~k=0,1,\dots,K, \label{eq:LiangPUSNRconstraint}\\
		~& \gamma_{s2} \geq \Gamma_s, \label{eq:s2sSNR}\\ 
		~&|\phi_n| = 1,~ n=1,\dots,N_r, \label{eq:LiangRISUniMod}
	\end{align}
\end{subequations}
where $\gamma_{s2}=\frac{T}{\delta^2}\Ebb_{c_{\ell}}[{|\g_0^H \bPhi^H \G^H\w c_{\ell}|^2}]$, $\Lambda_s = Q(\sqrt{2\Gamma_s})$ \cite{ShuRISSRTVT2025}, and $Q(x)$ denotes the $Q$-function. All results are averaged over $300$ independent channel realizations.

\begin{figure}[t!]
	\begin{center}
		{\resizebox{0.38\textwidth}{!}
			{\includegraphics{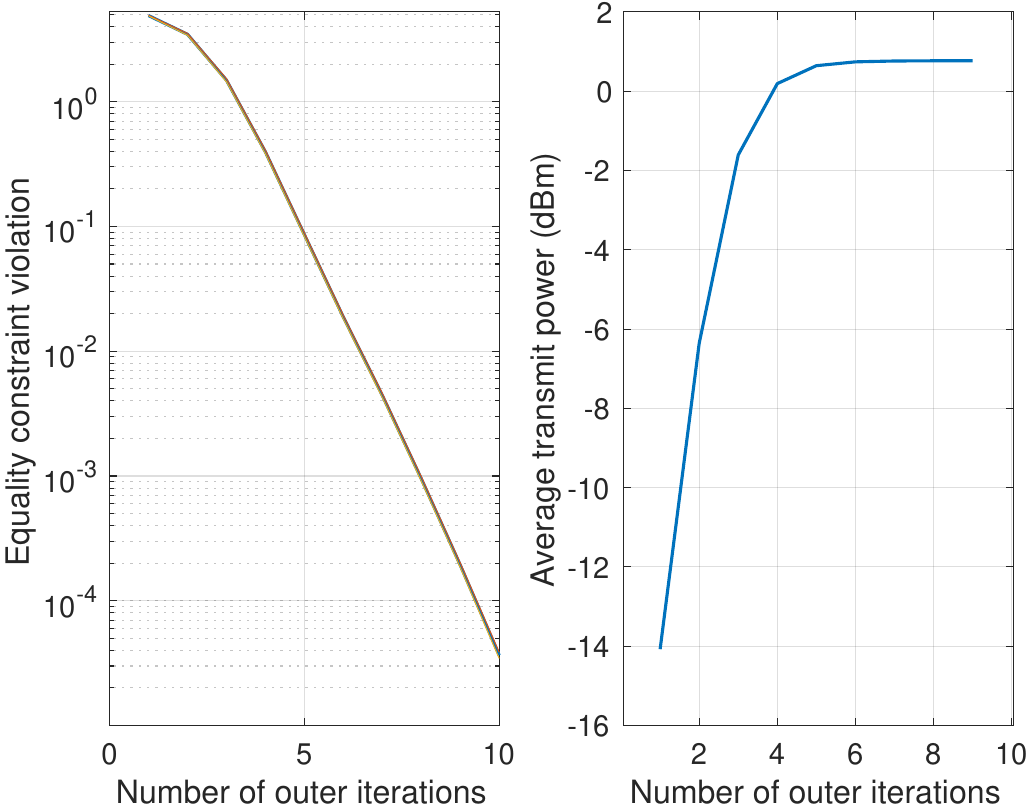}}}
	\end{center}\vspace{-0.5cm}
	\caption{Convergence behavior of Algorithm \ref{alg:BCD}. Left: equality constraint violation versus iterations; right: transmit power versus iterations.}\label{fig:convge}
	\vspace{-0.3cm}
\end{figure}
In the right panel of Fig.~\ref{fig:convge}, the transmit power increases monotonically as the penalty parameter $\rho$ grows {(since $\rho \leftarrow c_{\rho}\rho$ at each outer iteration as in Algorithm~\ref{alg:BCD})}, and converges within an average of $10$ iterations. Similarly, the equality constraint residual decreases linearly and reaches below $\epsilon_2 = 10^{-4}$ within $10$ iterations.

\begin{figure}[t!]
	\begin{center}
		{\resizebox{0.38\textwidth}{!}
			{\includegraphics{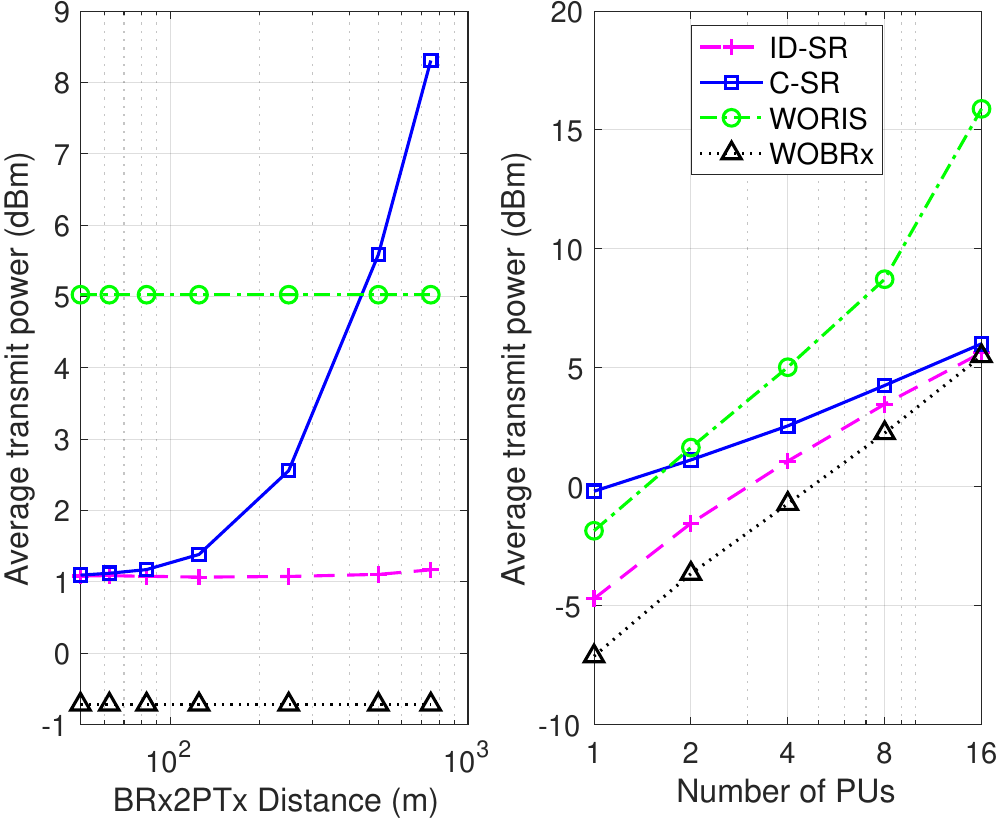}}}
	\end{center}\vspace{-0.5cm}
	\caption{Average transmit power versus spatial deployment and number of PRs. }\label{fig:Perfs}
	\vspace{-0.3cm}
\end{figure}
In the left panel of Fig.~\ref{fig:Perfs}, the RIS reduces the required power across deployments. The proposed ID-SR scheme consistently requires less power than C-SR. 
The C-SR's power increases quickly due to its implicit requirement to keep the BRx's primary link strong for joint decoding. 
On the right, the power of all RIS-enhanced systems converges to a common point as $K$ approaches $N_t$, indicating diminishing marginal cost as additional PRs share the same broadcast stream. 

\section{Conclusion}
In this paper, we studied a RIS-enhanced ID-SR architecture for broadcasting, where a multi-antenna PTx delivers a common stream to multiple PRs and a RIS-BDx conveys a secondary stream to a BRx with an energy detector. To minimize the transmit power under a common rate requirement for all PRs together with a BER constraint at the BRx, we 
adopted a rate-balanced reformulation and a monotonic BER ratio characterization, and 
developed a PBCD algorithm with closed-form updates. 
%
%
Simulations verified fast decay of the equality constraint violation and demonstrated consistent transmit power savings relative to conventional SR with joint decoding. 
Future work includes robust and stochastic designs under imperfect CSI and discrete-phase RIS constraints.
\bibliographystyle{IEEEbib}
\bibliography{bibs/SR}

\end{document}